\documentclass[fleqn,twoside]{article}
\usepackage{espcrc2}

\usepackage{graphicx}
\usepackage[figuresright]{rotating}

\title{Progress on a canonical finite density algorithm}

\author{
Andrei Alexandru 
\address[KY]
{Department of Physics and Astronomy, University of Kentucky, 
Lexington, KY 40506, USA} 
\thanks{Talks presented by Andrei Alexandru and Keh-Fei Liu at
Lattice 2004 conference.}, 
Manfried Faber\address[GE]
{Atomic Institute of the Austrian Universities, Nuclear Physics Division,
A-1040 Vienna, Austria},
Ivan Horv\'{a}th\addressmark[KY], 
Keh-Fei Liu\addressmark[KY]}

\begin{document}

\begin{abstract}
We test the finite density algorithm in the canonical ensemble which
combines the HMC update with the accept/reject step according to the ratio of 
the fermion number projected determinant to the unprojected one 
as a way of avoiding the determinant fluctuation problem.
We report our preliminary results on the Polyakov loop in different baryon 
number sectors which exhibit deconfinement transitions on small lattices.
The largest density we obtain around $T_c$ is an order of magnitude larger than that of
nuclear matter. From the conserved vector current, we calculate the quark number 
and verify that the mixing of different baryon sectors is small.

\end{abstract}

\maketitle

\section{Motivation}

The phase structure of QCD is relevant in the description of various
phenomena: from subtle modifications of the cross-sections in high
energy collisions of nuclei to exotic states of nuclear matter
in neutron stars. 
Even though we have 
a good understanding of how to perform simulations at zero baryon density and finite 
temperature, the simulation with  non-zero chemical potential has been
problematic.

To deal with the complex phase that the chemical potential introduces to the
fermion determinant, a number of techniques have been proposed:
various reweighting methods, imaginary chemical potential, etc. They
allow us to explore the phase structure at small values of the chemical
potential and temperatures close to the critical temperature. All
these methods have as a starting point the grand canonical partition
function. We will present here an approach based on the canonical
partition function which alleviates the sign problem, the overlap 
problem and the determinant fluctuation problems~\cite{liu03,liu02}.
Some preliminary results will also be reported.

\section{Canonical partition function}

The QCD canonical partition function with a fixed quark number can be
constructed from the projection of the quark number from the
determinant~\cite{fab95,liu02}.  The probability of having $k$ quark
loops more than antiquark loops wrapping around the time direction in
the loop expansion of a determinant is represented by the Fourier
transform
\begin{equation}  \label{projection}
P_{k} = \frac{1}{2 \pi} \int_0^{2 \pi} \mbox{d}\phi \, e^{-i \phi k} \det M_\phi.
\end{equation}
where $\det M_{\phi}$ is the fermion determinant of the quark matrix with an 
additional U(1) phase $e^{i\phi}/e^{-i\phi}$ in the time forward/backward 
links on one time slice. Therefore, the canonical partition function can be 
written as
\begin{equation}   \label{zc}
Z_C(k, T) = \int \mbox{\cal D}U \,
e^{-S_G} \int_0^{2\pi} \frac{\mbox{d}\phi}{2\pi}e^{-i\phi k} \det M_{\phi}.
\end{equation}

One can also derive this from the Fourier transform
of the grand canonical partition function~\cite{dms90}
$$ 
Z_C(k, T) = \int_0^{2\pi}\frac{\mbox{d}\phi}{2\pi} \, e^{-i\phi k}
Z_{GC}(\mu_q = i \phi T,T)
$$ 
To see this we can use the fugacity expansion of the grand
canonical partition function
\begin{equation}
Z_{GC}(\mu_q, T)=\sum_{k=-12V}^{12V} Z_C(k, T) e^{\mu_q k/T} 
\label{fe}
\end{equation}
where $k=n_q-n_{\bar{q}}$ is the quark number and $\mu_q$ is the quark chemical potential.
Using the usual expression for Lattice QCD grand canonical partition 
function, one obtains the same expression as in Eq. (\ref{zc}).


For Wilson fermions we can write
$ M_\phi(U) \equiv M(U_\phi) = M(\mu = i\phi T, U) $
in terms of the standard fermion matrix
\begin{eqnarray*}
[M(U)]_{m,n} &=& \delta_{m,n}-\kappa (1-\gamma_\mu)U_\mu(m)\delta_{m+\mu,n} \\
&-&\kappa (1+\gamma_\mu)U^\dagger_\mu(n)\delta_{m,n+\mu}
\end{eqnarray*}
and some modified gauge fields
\begin{eqnarray}
(U_\phi)_\mu(n) \equiv \left\{ 
\begin{array}{ll}
U_\mu(n) e^{-i\phi} & n_4=N_t ,\, \mu=4 \\
U_\mu(n) & otherwise .
\end{array}
\right.
\label{uf}
\end{eqnarray}
This should not be viewed as a change of the field variables but
rather as a convenient way of writing the partition function. We
should also note that although originally the phase $e^{-i\phi T}$ appears
on all timelike links, we can accumulate it on the last time slice
through a change of variables.

In order to evaluate the canonical partition function in Eq. (\ref{zc}) we
need to replace the continuous Fourier transform with a discrete
one. Using the discrete Fourier transform of the determinant
$$
\mbox{det}_k M(U) \equiv 
\frac{1}{N}\sum_{n=0}^{N-1} e^{-i k \phi_n} \det M_{\phi_n}(U) 
$$
with $\phi_n = \frac{2 \pi n}{N}$, we write 
the partition function as
\begin{equation}
\tilde{Z}_C(k)=\int\mbox{\cal D}U\, e^{-S_G(U)} \mbox{det}_k M(U)
\label{zcd}
\end{equation}
This is the partition function that we will try to evaluate. The
parameter $N$ defines the Fourier transform. In the limit
$N\rightarrow\infty$ we recover the original partition function. For
finite $N$ the partition function will only be an approximation of the
canonical partition function. Using the fugacity expansion we can show
that
$$
\tilde{Z}_C(k) = \sum_{m=-\infty}^{\infty} Z_C(k+m N)
$$
If $F_B$ denotes the baryon free energy we expect that
$$
Z_C(N_B)\propto e^{-|N_B| F_B/T},
$$
in the confined phase. If this holds true then $\tilde{Z}_C(k)\approx Z_C(k)$ 
for $k\ll N$. However, this is an expectation that needs to be checked in our 
simulations.

\section{Algorithm}

To evaluate $\tilde{Z}_C(k)$ by Monte-Carlo techniques we need the integrand in 
(\ref{zcd}) to be real and positive. Using the $\gamma_5$ hermiticity of
the Wilson matrix
it is easy to prove that $\det M(U_\phi)$
is real. However, the integrand in (\ref{zcd}) is the Fourier transform of
the Wilson determinant. For $\mbox{det}_k M$ to be real we need that
$\det M_\phi = \det M_{-\phi}$. From charge conjugation symmetry we know
that $\langle \det M_\phi \rangle = \langle \det M_{-\phi} \rangle$. 
Unfortunately this relation doesn't hold configuration by configuration. 
We are thus forced to remove the complex phase from $\mbox{det}_k M$.

To avoid dealing with the complex phase we will generate an ensemble using
$|\mbox{Re}\,\mbox{det}_k M|$ as our probability function and we will then
reintroduce the complex phase in observables. This is the standard way
to deal with a complex phase in the weight function. It is also the source
of the infamous sign problem. This might limit us in fully exploring the parameter
space.

The focus of the algorithm will then be to generate an ensemble with the 
weight
$$
W(U) = e^{-S_G(U)}
\left| \frac{1}{N}\sum_{n=0}^{N-1} \cos(\phi_n k) \det M_{\phi_n}(U)\right|.
$$

It was pointed out in~\cite{alf99} that taking the Fourier transform after the
Monte Carlo simulations with different $\phi_n$ leads to an overlap problem.
To avoid this problem, it was stressed~\cite{liu02,liu03} that it is essential
to perform the Fourier transform first to lock into one particular quark (or
baryon) sector before the accept/reject step. The general approach is to
break the updating process into two steps. In this case, one first uses an update 
for an approximate weight $W'(U)$, and then use an accept/reject step to correct
for the approximation. The efficiency of this strategy depends on how good the
approximation is. For a poor approximation the acceptance rate in the
second step will be very small. 

One possible solution would be to update the gauge links using a
heatbath for pure gauge action. In that case
$W'(U)=e^{-S_G(U)}$. However, it is known that such an updating 
strategy is very inefficient since the fermionic part is completely
disregarded in the proposal step and the determinant, being an extensive quantity,
can fluctuate wildly from one configuration to the next in the pure gauge 
updating process \cite{jhl03}. To avoid the overlap problem and the leading 
fluctuation in the determinant, it is proposed~\cite{liu03} 
to update using the HMC algorithm in the first step. For simplicity we will 
be simulating two degenerate flavors of quarks. This allows us to use the standard 
HMC update. In this case we have
$$
W'(U)=e^{-S_G(U)} \det M(U)
$$

In the second step we have to accept/reject with the probability
$$
P_{acc} = \min \left\{1, \omega(U')/\omega(U) \right\} 
$$
where $\omega$ is the ratio of the weights
$$
\omega(U)= \frac{W(U)}{W'(U)}=
\frac{|\mbox{Re}\,\mbox{det}_k M[U]|}{\det M[U]}.
$$
Since $\mbox{det}_k M[U]$ and $\det M[U]$ are calculated on the same
gauge configuration, we can write the determinant ratio as
\begin{equation}
\omega(U)
= \frac{1}{N} \left| \sum_{n=0}^{N-1} \cos(\phi_n k) 
e^{Tr(\log M_{\phi_n} - \log M)}\right|.
\end{equation}
The leading  fluctuation in the determinant from one gauge
configuration to the next is removed by the $Tr \log$ difference of the
quark matrices of $M_{\phi_n}(U)$ and $M(U)$. This should 
greatly improve the acceptance rate compared to the case with direct 
accept/reject based on the $|\mbox{Re}\,\mbox{det}_k M(U)|$ itself.

\section{Triality}

The canonical partition function (\ref{zc}) has a $Z_3$ symmetry that is not
present in the grand canonical partition function~\cite{fbm95}. Under a
transformation $U\rightarrow U_\phi$ where $U_\phi$ is defined as in (\ref{uf})
and $\phi = \pm \frac{2\pi}{3}$ we have
$$
\mbox{det}_k M[U]\rightarrow \mbox{det}_k M[U_{\pm \frac{2\pi}{3}}]=
e^{\pm i\frac{2\pi}{3}k}\mbox{det}_k M[U].
$$
We see that when $k$ is a multiple of $3$ this transformation is a
symmetry of the action. Incidentally, if this symmetry is not
spontaneously broken the relation above guarantees that the canonical
partition function will vanish for any $k$ that is not a multiple of $3$.

For $\tilde{Z}_C$, our approximation of the canonical partition function,
to exhibit this symmetry we need to choose the parameter $N$ that
defines the discrete Fourier transform to be a multiple of $3$. 

\begin{figure}
\includegraphics[height=1.8in]{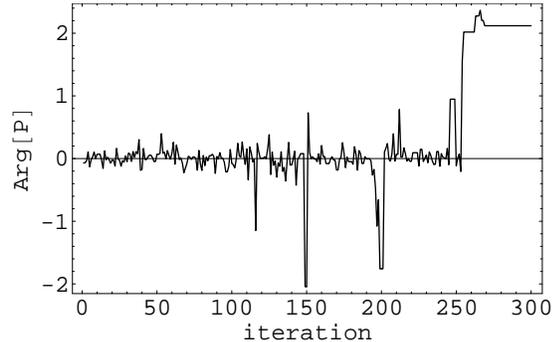}
\caption{Polyakov loop argument as a function of simulation time.
\label{freeze}}
\end{figure}

If N is a multiple of $3$ we have 
$$
W(U)=W(U_{\pm\frac{2\pi}{3}}).
$$
However, the HMC weight $W'$ does not respect this symmetry since $\det M$ is
not symmetric under this transformation. In this case, our algorithm can become
frozen for long periods of time. In Fig. \ref{freeze} we show the argument
of the Polyakov loop as a function of the simulation time. Under a $Z_3$ 
transformation $U\rightarrow U_{\pm\frac{2\pi}{3}}$ the argument of the 
Polyakov loop gets shifted by $\pm\frac{2\pi}{3}$. The HMC update
strongly prefers the $0$ sector and when a tunneling occurs to one of the
other sectors the accept/reject step will reject any change for a long period
of time. Consequently the algorithm will be very poor in sampling the other
sectors.

To alleviate this problem we will introduce a $Z_3$
hopping~\cite{kf03}. Since the weight $W$ is symmetric under this
change we can intermix the regular updates with a change in the field
variables $U\rightarrow U_{\pm\frac{2\pi}{3}}$, where we will choose
the sign randomly (equal probability for each sign) to satisfy the
detailed balance. This will decrease our acceptance by a factor of
roughly $3$ but will sample all the $Z_3$ sectors in the same manner.

\section{Simulation details}

In order to simulate the ensemble given by the weight $W(U)$ we need to 
evaluate the Fourier transform of the determinant. Since the calculation
of the determinant is quite time consuming we will need to use $N$ as small
as possible. There is a proposal that employs an estimator for the 
determinant~\cite{tdl98} but in this preliminary study we choose to 
calculate the determinant exactly using LU decomposition. We were thus 
limited to running on very small lattices, i.e. $4^4$, and with $N=12$. For
each $\beta$ we run three simulations: one for $k=0$, one for $k=3$ and
one for $k=6$. They correspond to $0$, $1$ and $2$ baryons in the box.

\begin{table}
\caption{The simulation parameters for our runs.\label{t1}}
\begin{tabular}{|c|c|c|c|c|}
\hline
$\beta$
&$a(\mbox{fm}$) & $m_\pi(\mbox{MeV})$ & $V^{-1}(\mbox{fm}^{-3})$ & 
$T(\mbox{MeV})$\\
\hline
5.1 & 0.35(3) &  870(90)  & 0.36(12) & 140(14) \\
5.2 & 0.31(2) &  920(90)  & 0.52(17) & 160(16) \\
5.3 & 0.24(2) & 1050(100) & 1.1(3)   & 205(20) \\
\hline
\end{tabular}
\end{table}

In order to get a reasonable box size we had to simulate at large lattice 
spacings. We used $\beta=5.1$, $5.2$ and $5.3$ with fixed $\kappa = 0.158$.
The relevant parameters can be found in Table \ref{t1}. 
The lattice spacing is determined using standard dynamical action on a $12^4$ 
lattice for the same values of $\beta$ and $\kappa$.
For
the HMC part of the update we used trajectories of length $0.5$ with 
$\Delta\tau=0.01$. We adjusted the number of such trajectories between 
two consecutive accept/reject steps so that the acceptance stays within
the $15\%$ to $30\%$ range. The configurations were saved after $10$ accept/reject
steps. We collected about $100$ such configurations
for each run.

\section{Observables}

As we mentioned earlier we need to reintroduce the phase when we compute
the observables. For an observable $O$ we have the following expression
$$
\langle O\rangle_{\tilde{Z}_C} = \frac
{\langle O \alpha \rangle_W}{\langle \alpha\rangle_W}
$$
where the subscript $\tilde{Z}_C$ stands for integration over the canonical
ensemble given by (\ref{zcd}), $W$ stands for the average over the weight
$W$ and
$$
\alpha \equiv \frac{\mbox{det}_k M}{|\mbox{Re}\,\mbox{det}_k M|}
$$
is the reweighting phase. It is easy to see that the phase $\alpha$
should average, over an infinite ensemble, to a real number. This is
direct consequence of the charge conjugation symmetry. The real part
of this phase will always be $\pm1$. This gives us a simple criterion
to decide whether we have a sign problem; all we need to do is to check
whether the phase has predominantly one sign. If the phase comes with
plus and minus signs with close to equal probability then we have a sign
problem. For our runs the data is 
 collected in Table \ref{sign}. We
see that for $\beta=5.1$ and $k=6$ we are approaching the sign problem
region.

\begin{table}
\caption{Average of real phase factor, $\langle\mbox{Re}\,\alpha\rangle_W$. 
\label{sign}}
\begin{tabular}{|c|c|c|c|}
\hline
&k=0 & k=3 & k=6 \\
\hline
$\beta$=5.1 & 1.00(0) & 0.48(6) & 0.17(10) \\
$\beta$=5.2 & 1.00(0) & 0.61(8) & 0.82(5)  \\
$\beta$=5.3 & 1.00(0) & 0.98(1) & 0.96(3) \\
\hline
\end{tabular}
\end{table}

\begin{figure}
\includegraphics[height=1.8in]{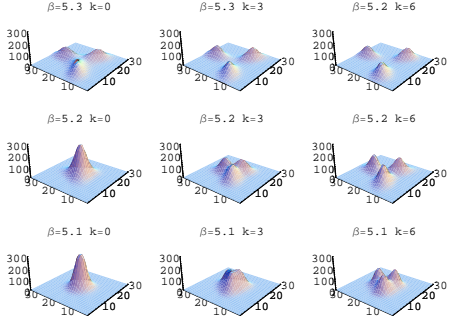}
\caption{Polyakov loop distribution. \label{poly}}
\end{figure}

An interesting observable to measure is the Polyakov loop. Although the average 
value is expected to be zero due to the $Z_3$ symmetry, the  distribution
in the complex plane can tell us whether or not the symmetry is spontaneously broken.
Usually the spontaneous breaking of the $Z_3$ symmetry is associated with 
deconfinement. In Fig. \ref{poly} we show the histograms of the Polyakov loop 
for our runs. We see that as we go from $\beta=5.2$ to $\beta=5.3$ the 
symmetry is broken even in the $k=0$ sector. This is the usual finite 
temperature phase transition at zero baryon density. According to 
Table~\ref{t1} this happens somewhere between $160\,\mbox{MeV}$ and 
$205\,\mbox{MeV}$. This is somewhat higher than expected for dynamical 
simulations but can be explained in terms of our large quark masses. 
More interestingly, we see that under
the critical temperature the transition still occurs as we increase the
number of baryons. It is also apparent that as we lower $\beta$ ($\beta = 5.1$), 
we are approaching a region where even the $N_B=1$ sector is confined. 

Assuming that the result of $N_B = 1 (k =3)$ at $\beta = 5.1$ is close to
the phase transition line, the phase transition for $T = 140\,\mbox{MeV}$
occurs at the baryon density of $0.36\,\mbox{fm}^{-3}$ which is about $2.3$ times the
nuclear matter density. We see from Table 1 that there does not seem to be a problem 
of reaching a density an order of magnitude higher than that of the nuclear matter
around $T_c$. For example, for $N_B = 2 (k =6)$ at $\beta = 5.3$, the baryon
density is $2.2(6)\,\mbox{fm}^{-3}$ which is $14$ times the nuclear matter density.
However, we should not take these numbers seriously in view of the
small volume and large quark mass in the present calculation 
(since the pion mass is $870\,\mbox{MeV}$ our quark mass is close to the 
strange quark one).

While it is nice to see that our expectation for the phase transition are
confirmed in the gluonic sector, one would like to explore this
transition in the fermionic sector too. It is important to note that the
fermionic observables have to be evaluated in a non-standard manner.
For example, if we are to study the fermionic bilinears we get
\begin{eqnarray*}
\langle \bar{\psi}\Gamma\psi\rangle &=& \frac{1}{Z_C}\frac{1}{N}
\sum_{\phi} e^{-ik\phi}\int \mbox{\cal D}\bar{\psi}\mbox{\cal D}\psi
e^{\bar{\psi}M_{\phi}\psi} \bar{\psi}\Gamma\psi \\
&=& \sum_{k'}\frac{\mbox{det}_{k'} M}{\mbox{det}_{k} M} (-\mbox{Tr}_{k-k'}
\Gamma M^{-1}) 
\end{eqnarray*}
where
$$
\mbox{Tr}_k \Gamma M^{-1} = \frac{1}{N}\sum_{\phi_n} e^{-ik\phi_n}
\mbox{Tr}\Gamma M_{\phi_n}^{-1}.
$$
We see that for fermionic variables, we need to do a Fourier transform
on the observable too. We have measured the chiral condensate in our runs
in the hope that it will signal chiral symmetry restoration. Unfortunately,
it seems that the quarks are too heavy for us to get any signal. For example
at $\beta=5.1$ and $\kappa=0.158$ we got 
$\langle\bar{\psi}\psi\rangle=-3.612(1)$, $-3.598(5)$ and $-3.601(2)$ for
$N_B=0$, $1$ and $2$ respectively. We suspect that we need to lower the quark 
mass to be able to see any signal in the chiral condensate. Also, since we are working with
Wilson fermions, the signal might be obscured by the mixing between 
$\bar{\psi}\psi$ and unity and other lattice artifacts
especially at these large lattice spacings.

\section{Sector mixing}

It is important to note that all the calculations carried out above are
using an approximation of the canonical partition function. As noted before
it is an assumption on our part that $\tilde{Z}_C\approx Z_C$. To check this
we measured the conserved charge for Wilson fermions
\begin{eqnarray*}
Q(t)=-\kappa \sum_{\vec{x}}&[\bar{\psi}(x) U_4(x)(1-\gamma_4)
\psi(x+\hat{t}) \\
& -\bar{\psi}(x+\hat{t})U^\dagger_4(x)(1+\gamma_4)\psi(x)].
\end{eqnarray*}
The idea is that if we were to compute the conserved charge using the 
exact canonical ensemble $Z_C$ we would measure exactly the number of 
quarks we put it $\langle Q(t)\rangle_{Z_C(k)}=k$. However, in the discrete
case we have
$$
\langle Q(t) \rangle_{\tilde{Z}_C(k)} = \frac{\sum_m (k+m N)Z_C(k+mN)}
{\sum_m Z_C(k+mN)}.
$$ 
This is a weighted average over the various sectors allowed by our discretized
delta function. If the approximation holds then the average of the conserved
current should be closed to its exact value. 

For $\beta=5.1$, we get the conserved charge to be $\langle Q(t)
\rangle_{\tilde{Z}_C}=3.0004(4)$ for $k = 3$ and $0$ for $k=0$ and
$6$. We will first note that for $k=3$ we get the conserved charge to
be very close to the exact value, indicating that there is very little
mixing with the other allowed sectors $k=...-21,-9,15,27...$.  The
other two values are $0$ due to symmetry. For $k=0$ sector, it is easy to
understand. For the $k=6$ sector, the explanation lies in the fact that
$N=12$.  Thus for every allowed sector $k+mN$ there is an $m'$ such
that $k+m'N=-(k+mN)$. In that case the sum in the numerator vanishes
and the conserved charge ends up being zero.

We conclude that there is very little mixing with the other allowed
sectors in our runs. We attribute the smallness of the deviation in the
case $k=3$ to the fact that the free energy for our baryons is very
large. However, we expect that as we lower the quark mass the
deviation will become more significant. If the deviation is too large
then we might have to increase the parameter $N$ for the discrete
Fourier transform.

\section{Conclusions and outlook}

Summing up, we showed that the canonical partition function approach
can be used to investigate the phase structure of QCD. 
We are able to
explore temperatures around the critical temperature and much higher
densities than those available by other methods. We also find evidence
that sign fluctuations might limit us in reaching much lower temperature and
large fermion number.
We checked that the discrete Fourier approximation to the canonical
partition function introduces only minimal deviations.

In our runs, we find that even for a small lattice, the picture that
emerges for the phase structure is consistent with expectations. The
Polyakov loop shows signals of deconfining phase transition both in
temperature and density directions. To complement this picture we
would like to see a signal in the fermionic sector. For this we need
to go to smaller masses and, maybe, to finer lattices.

Another goal would be to locate certain points on the phase transition
line.  For this we need to get to even lower temperatures and/or
densities. While, we might be limited in going to much lower temperatures, we
should be able to reach lower densities. 


To reach a lower density with larger volume, we need to
introduce an esitmator for the determinant.  We should stress that the
method used to generate the modified ensemble has no bearing on
whether we have a sign problem or not. It is an intrinsic property of
the modified ensemble. Consequently, we do not expect any difference
in the sign behavior as we introduce the estimator.


\end{document}